\pdfoutput=1
\newif\ifarxiv
\arxivtrue
\ifarxiv\pdfmapfile{+classico.map}\fi
\newif\ifafour
\afourfalse% true = A4, false = A5
\newif\iftypodisclaim % typographical disclaim on the side
\typodisclaimtrue

\documentclass[\ifafour a4paper,12pt,\else a5paper,10pt,\fi%extrafontsizes,%
onecolumn,oneside,article,%french,italian,german,swedish,latin,
british%
]{memoir}

\newcommand*{\firstpublished}{13 July 2020}
\newcommand*{\updated}{\ifarxiv 31 July 2020\else\today\fi}
\newcommand*{\propertitle}{The rule of conditional probability\\ is valid in quantum theory%\\{\large ***}%
}% title uses LARGE; set Large for smaller
\newcommand*{\pdftitle}{The rule of conditional probability is valid in quantum theory}
\newcommand*{\headtitle}{Conditional probability is valid in quantum theory}
\newcommand*{\pdfauthor}{P.G.L.  Porta Mana}
\newcommand*{\headauthor}{Porta Mana}
\newcommand*{\reporthead}{\ifarxiv\else Open Science Framework \href{https://doi.org/10.31219/osf.io/bsnh7}{\textsc{doi}:10.31219/osf.io/bsnh7}\fi}% Report number

\usepackage[T1]{fontenc} 
\input{glyphtounicode} \pdfgentounicode=1

\usepackage[utf8]{inputenx}

\usepackage{textcomp}

\usepackage{amsmath}

\usepackage{mathtools}
\usepackage{amssymb}

\usepackage{amsxtra}

\usepackage[main=british,french,italian,german,swedish,latin,esperanto]{babel}

\usepackage[autostyle=false,autopunct=false,english=british]{csquotes}
\setquotestyle{american}
\newcommand*{\defquote}[1]{`\,#1\,'}

\usepackage{amsthm}

\theoremstyle{remark}

\newtheoremstyle{innote}{\parsep}{\parsep}{\footnotesize}{}{}{}{0pt}{}
\theoremstyle{innote}

\usepackage[shortlabels,inline]{enumitem}
\SetEnumitemKey{para}{itemindent=\parindent,leftmargin=0pt,listparindent=\parindent,parsep=0pt,itemsep=\topsep}
\setlist{itemsep=0pt,topsep=\parsep}
\setlist[enumerate,2]{label=\alph*.}
\setlist[enumerate]{label=\arabic*.,leftmargin=1.5\parindent}
\setlist[itemize]{leftmargin=1.5\parindent}
\setlist[description]{leftmargin=1.5\parindent}
\usepackage[babel,theoremfont,largesc]{newpxtext}

\usepackage[bigdelims,nosymbolsc%,smallerops % probably arXiv doesn't have it
]{newpxmath}
\makeatletter
\def\re@DeclareMathSymbol#1#2#3#4{%
    \let#1=\undefined
    \DeclareMathSymbol{#1}{#2}{#3}{#4}}
\re@DeclareMathSymbol{\bigoplusop}{\mathop}{largesymbols}{"4C}
\re@DeclareMathSymbol{\bigotimesop}{\mathop}{largesymbols}{"4E}
\re@DeclareMathSymbol{\sumop}{\mathop}{largesymbols}{"50}
\re@DeclareMathSymbol{\prodop}{\mathop}{largesymbols}{"51}
\re@DeclareMathSymbol{\bigcupop}{\mathop}{largesymbols}{"53}
\re@DeclareMathSymbol{\bigcapop}{\mathop}{largesymbols}{"54}
\re@DeclareMathSymbol{\bigwedgeop}{\mathop}{largesymbols}{"56}
\re@DeclareMathSymbol{\bigveeop}{\mathop}{largesymbols}{"57}
\re@DeclareMathSymbol{\bigtimesop}{\mathop}{largesymbolsPXA}{"10}
\makeatother
\DeclareFontFamily{U}{egreek}{\skewchar\font'177}%
\DeclareFontShape{U}{egreek}{m}{n}{<-6>s*[1]eurm5 <6-8>s*[1]eurm7 <8->s*[1]eurm10}{}%
\DeclareFontShape{U}{egreek}{m}{it}{<->s*[1]eurmo10}{}%
\DeclareFontShape{U}{egreek}{b}{n}{<-6>s*[1]eurb5 <6-8>s*[1]eurb7 <8->s*[1]eurb10}{}%
\DeclareFontShape{U}{egreek}{b}{it}{<->s*[1]eurbo10}{}%
\DeclareSymbolFont{egreeki}{U}{egreek}{m}{it}%
\SetSymbolFont{egreeki}{bold}{U}{egreek}{b}{it}% from the amsfonts package
\DeclareSymbolFont{egreekr}{U}{egreek}{m}{n}%
\SetSymbolFont{egreekr}{bold}{U}{egreek}{b}{n}% from the amsfonts package
\DeclareFontFamily{U}{egreekx}{\skewchar\font'177}
\DeclareFontShape{U}{egreekx}{m}{n}{%
       <-7.5>s*[0.9]euex7%
    <7.5-8.5>s*[0.9]euex8%
    <8.5-9.5>s*[0.9]euex9%
    <9.5->s*[0.9]euex10%
}{}
\DeclareSymbolFont{egreekx}{U}{egreekx}{m}{n}
\DeclareMathSymbol{\sumop}{\mathop}{egreekx}{"50}
\DeclareMathSymbol{\prodop}{\mathop}{egreekx}{"51}
\DeclareMathSymbol{\coprodop}{\mathop}{egreekx}{"60}
\makeatletter
\def\sum{\DOTSI\sumop\slimits@}
\def\prod{\DOTSI\prodop\slimits@}
\def\coprod{\DOTSI\coprodop\slimits@}
\makeatother
\renewcommand\sfdefault{uop}
\DeclareMathAlphabet{\mathsf}  {T1}{\sfdefault}{m}{sl}
\SetMathAlphabet{\mathsf}{bold}{T1}{\sfdefault}{b}{sl}
\newcommand*{\mathte}[1]{\textbf{\textit{\textsf{#1}}}}
\usepackage[scaled=0.84]{DejaVuSansMono}

\usepackage{mathdots}

\usepackage[usenames]{xcolor}
\definecolor{mypurpleblue}{RGB}{68,119,170}
\definecolor{myblue}{RGB}{102,204,238}
\definecolor{mygreen}{RGB}{34,136,51}
\definecolor{myyellow}{RGB}{204,187,68}
\definecolor{myred}{RGB}{238,102,119}
\definecolor{myredpurple}{RGB}{170,51,119}
\definecolor{mygrey}{RGB}{187,187,187}
\definecolor{lgrey}{RGB}{221,221,221}
\colorlet{shadecolor}{lgrey}

\usepackage{bm}

\usepackage{microtype}

\usepackage[backend=biber,mcite,%subentry,
citestyle=authoryear-comp,bibstyle=pglpm-authoryear,autopunct=false,sorting=ny,sortcites=false,natbib=false,maxcitenames=2,maxbibnames=8,minbibnames=8,giveninits=true,uniquename=false,uniquelist=false,maxalphanames=1,block=space,hyperref=true,defernumbers=false,useprefix=true,sortupper=false,language=british,parentracker=false]{biblatex}
\DeclareSortingScheme{ny}{\sort{\field{sortname}\field{author}\field{editor}}\sort{\field{year}}}
\DefineBibliographyExtras{british}{\def\finalandcomma{\addcomma}}
\renewcommand*{\finalnamedelim}{\addspace\amp\space}
\DeclareDelimFormat{multicitedelim}{\addsemicolon\addspace\space}
\DeclareDelimFormat{compcitedelim}{\addsemicolon\addspace\space}
\DeclareDelimFormat{postnotedelim}{\addspace}
\ifarxiv\else\fi
\renewcommand{\bibfont}{\footnotesize}
\defbibheading{bibliography}[\bibname]{\section*{#1}\addcontentsline{toc}{section}{#1}%\markboth{#1}{#1}
}
\newcommand*{\citep}{\parencites}%{\footcites}
\newcommand*{\citey}{\parencites*}
\newcommand*{\ibid}{\unspace\addtocounter{footnote}{-1}\footnotemark{}}
\providecommand{\href}[2]{#2}

\newcommand*{\amp}{\&}

\newcommand*{\subtitleproc}[1]{}

\usepackage{graphicx}

\PassOptionsToPackage{hyphens}{url}\usepackage[hypertexnames=false]{hyperref}

\usepackage[depth=4]{bookmark}
\hypersetup{colorlinks=true,bookmarksnumbered,pdfborder={0 0 0.25},citebordercolor={0.2667 0.4667 0.6667},citecolor=mypurpleblue,linkbordercolor={0.6667 0.2 0.4667},linkcolor=myredpurple,urlbordercolor={0.1333 0.5333 0.2},urlcolor=mygreen,breaklinks=true,pdftitle={\pdftitle},pdfauthor={\pdfauthor}}

\usepackage[british]{datetime2}
\DTMnewdatestyle{mydate}%
{% definitions
}
\DTMsetdatestyle{mydate}

\ifafour\setstocksize{297mm}{210mm}%{*}% A4
\else\setstocksize{210mm}{5.5in}%{*}% 210x139.7
\fi
\settrimmedsize{\stockheight}{\stockwidth}{*}
\setlxvchars[\normalfont] %313.3632pt for a 66-characters line
\setxlvchars[\normalfont]
\ifafour\settypeblocksize{*}{32pc}{1.618} % A4
\else\settypeblocksize{*}{26pc}{1.618}% nearer to a 66-line newpx and preserves GR
\fi
\setulmargins{*}{*}{1}%gives equal margins
\setlrmargins{*}{*}{*}
\setheadfoot{\onelineskip}{2.5\onelineskip}
\setheaderspaces{*}{2\onelineskip}{*}
\setmarginnotes{2ex}{10mm}{0pt}
\checkandfixthelayout[nearest]
\fixpdflayout
\pdfmapline{+dummy-space <dummy-space.pfb}\pdfinterwordspaceon%

\newenvironment{acknowledgements}{\section*{Thanks}\addcontentsline{toc}{section}{Thanks}}{\par}
\makeatletter\renewcommand{\appendix}{\par
  \bigskip{\centering
   \interlinepenalty \@M
   \normalfont
   \printchaptertitle{\sffamily\appendixpagename}\par}
  \setcounter{section}{0}%
  \gdef\@chapapp{\appendixname}%
  \gdef\thesection{\@Alph\c@section}%
  \anappendixtrue}\makeatother
\counterwithout{section}{chapter}
\setsecnumformat{\upshape\csname the#1\endcsname\quad}
\setsecheadstyle{\large\bfseries\sffamily%
\centering}
\setsubsecheadstyle{\bfseries\sffamily%
\raggedright}
\setsubsecindent{0pt}%0ex plus 1ex minus .2ex}
\setparaheadstyle{\bfseries\sffamily%
\raggedright}
\newcommand{\addchap}[1]{\chapter*[#1]{#1}\addcontentsline{toc}{chapter}{#1}}
\newcommand{\addsec}[1]{\section*{#1}\addcontentsline{toc}{section}{#1}}

\copypagestyle{manaart}{plain}
\makeheadrule{manaart}{\headwidth}{0.5\normalrulethickness}
\makeoddhead{manaart}{%
{\footnotesize%\sffamily%
\scshape\headauthor}}{}{{\footnotesize\sffamily%
\headtitle}}
\makeoddfoot{manaart}{}{\thepage}{}
\newcommand*\autanet{\includegraphics[height=\heightof{M}]{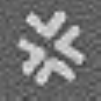}}
\definecolor{mygray}{gray}{0.333}
\iftypodisclaim%
\ifafour\newcommand\addprintnote{\begin{picture}(0,0)%
\put(245,149){\makebox(0,0){\rotatebox{90}{\tiny\color{mygray}\textsf{This
            document is designed for screen reading and
            two-up printing on A4 or Letter paper}}}}%
\end{picture}}% A4
\else\newcommand\addprintnote{\begin{picture}(0,0)%
\put(176,112){\makebox(0,0){\rotatebox{90}{\tiny\color{mygray}\textsf{This
            document is designed for screen reading and
            two-up printing on A4 or Letter paper}}}}%
\end{picture}}\fi%afourtrue
\makeoddfoot{plain}{}{\makebox[0pt]{\thepage}\addprintnote}{}
\else
\makeoddfoot{plain}{}{\makebox[0pt]{\thepage}}{}
\fi%typodisclaimtrue
\makeoddhead{plain}{\scriptsize\reporthead}{}{}
\pretitle{\begin{center}\LARGE\sffamily%
\bfseries}
\posttitle{\bigskip\end{center}}

\makeatletter\newcommand*{\atf}{\includegraphics[%trim=1pt 1pt 0pt 0pt,
totalheight=\heightof{@}]{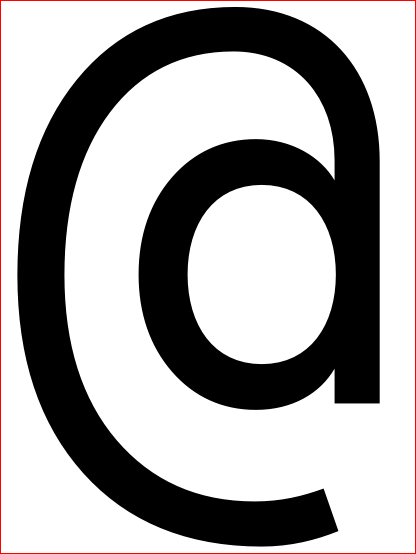}}\makeatother

\providecommand{\epost}[1]{\texttt{\footnotesize\textless#1\textgreater}}
\providecommand{\email}[2]{\href{mailto:#1ZZ@#2 ((remove ZZ))}{#1\protect\atf#2}}

\preauthor{\vspace{-0.5\baselineskip}\begin{center}
\normalsize\sffamily%
\lineskip  0.5em}
\postauthor{\par\end{center}}
\predate{\DTMsetdatestyle{mydate}\begin{center}\footnotesize}
\postdate{\end{center}\vspace{-\medskipamount}}

\setfloatadjustment{figure}{\footnotesize}
\captiondelim{\quad}
\captionnamefont{\footnotesize\sffamily%
}
\captiontitlefont{\footnotesize}
\midsloppy
\paragraphfootnotes
\footmarkstyle{\textsuperscript{%\color{myred}
\scriptsize\bfseries#1}~}
\title{\propertitle}
\author{%
\hspace*{\stretch{1}}%
\parbox{0.75\linewidth}%\makebox[0pt][c]%
{\protect\centering P.G.L.  Porta Mana  \href{https://orcid.org/0000-0002-6070-0784}{\protect\includegraphics[scale=0.16]{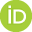}}\\%
\footnotesize Kavli Institute, Trondheim\quad\epost{\email{pgl}{portamana.org}}}%
\hspace*{\stretch{1}}%
}

\date{\firstpublished; updated \updated}

\newcommand*{\I}{\mathrm{i}}%imaginary unit
\newcommand*{\e}{\mathrm{e}}%Neper
\DeclareMathOperator{\tr}{tr}%trace
\DeclarePairedDelimiter\clcl{[}{]}
\DeclarePairedDelimiter\abs{\lvert}{\rvert}
\DeclarePairedDelimiter\set{\{}{\}}
\newcommand*{\pf}{\mathrm{p}}%probability
\newcommand*{\p}{\mathrm{P}}%probability
\renewcommand*{\|}[1][]{\nonscript\,#1\vert\nonscript\;\mathopen{}}
\newcommand*{\sect}{\S}% Sect.~
\newcommand*{\chap}{ch.}%
\newcommand*{\chaps}{chs}%
\newcommand*{\eqn}{eq.}%
\newcommand*{\eqns}{eqs}%

\newcommand*{\eg}{{e.g.}}

\newcommand*{\cf}{{cf.}}
\newcommand*{\etal}{{et al.}}

\newcommand*{\tsum}{\mathop{\textstyle\sum}\nolimits}
\definecolor{notecolour}{RGB}{68,170,153}

\newcommand*{\yDa}{\textcolor{mypurpleblue}{D_{\textrm{a}}}}
\newcommand*{\yDb}{\textcolor{myredpurple}{D_{\textrm{b}}}}
\newcommand*{\yDc}{\textcolor{mygreen}{D_{\textrm{c}}}}
\newcommand*{\yDd}{\textcolor{myyellow}{D_{\textrm{d}}}}
\newcommand*{\yr}{\bm{\rho}}
\newcommand*{\yM}{\mathte{M}}
\newcommand*{\yxa}{X_{1}}
\newcommand*{\yxb}{X_{2}}
\newcommand*{\yDq}{D_{q}}
\newcommand*{\yDqa}{\textcolor{mypurpleblue}{D_{q'}}}
\newcommand*{\yDqb}{\textcolor{myredpurple}{D_{q''}}}
\newcommand*{\yqa}{\textcolor{mypurpleblue}{q'}}
\newcommand*{\yqb}{\textcolor{myredpurple}{q''}}
\newcommand*{\povm}{\textsc{povm}}
\newcommand*{\ybr}{\bm{r}}
\begin{document}
\captiondelim{\quad}\captionnamefont{\footnotesize}\captiontitlefont{\footnotesize}
\selectlanguage{british}\frenchspacing
\maketitle

%%%%%%%%%%%%%%%%%%%%%%%%%%%%%%%%%%%%%%%%%%%%%%%%%%%%%%%%%%%%%%%%%%%%%%%%%%%%
%%% Abstract
%%%%%%%%%%%%%%%%%%%%%%%%%%%%%%%%%%%%%%%%%%%%%%%%%%%%%%%%%%%%%%%%%%%%%%%%%%%%
\iffalse\abstractrunin
\abslabeldelim{}
\renewcommand*{\abstractname}{}
\setlength{\absleftindent}{0pt}
\setlength{\absrightindent}{0pt}
\setlength{\abstitleskip}{-\absparindent}
\begin{abstract}\labelsep 0pt%
  \noindent ***
% \\\noindent\emph{\footnotesize Note: Dear Reader
%     \amp\ Peer, this manuscript is being peer-reviewed by you. Thank you.}
% \par%\\[\jot]
% \noindent
% {\footnotesize PACS: ***}\qquad%
% {\footnotesize MSC: ***}%
%\qquad{\footnotesize Keywords: ***}
\end{abstract}\fi
\selectlanguage{british}\frenchspacing

%%%%%%%%%%%%%%%%%%%%%%%%%%%%%%%%%%%%%%%%%%%%%%%%%%%%%%%%%%%%%%%%%%%%%%%%%%%%
%%% Epigraph
%%%%%%%%%%%%%%%%%%%%%%%%%%%%%%%%%%%%%%%%%%%%%%%%%%%%%%%%%%%%%%%%%%%%%%%%%%%%
% \asudedication{\small ***}
% \vspace{\bigskipamount}
% \setlength{\epigraphwidth}{.7\columnwidth}
% %\epigraphposition{flushright}
% \epigraphtextposition{flushright}
% %\epigraphsourceposition{flushright}
% \epigraphfontsize{\footnotesize}
% \setlength{\epigraphrule}{0pt}
% %\setlength{\beforeepigraphskip}{0pt}
% %\setlength{\afterepigraphskip}{0pt}
% \epigraph{\emph{text}}{source}

%%%%%%%%%%%%%%%%%%%%%%%%%%%%%%%%%%%%%%%%%%%%%%%%%%%%%%%%%%%%%%%%%%%%%%%%%%%%
%%% BEGINNING OF MAIN TEXT
%%%%%%%%%%%%%%%%%%%%%%%%%%%%%%%%%%%%%%%%%%%%%%%%%%%%%%%%%%%%%%%%%%%%%%%%%%%%

In a recent manuscript, Gelman \amp\ Yao \citey{gelmanetal2020} claim that
\enquote{the usual rules of conditional probability fail in the quantum
  realm} and that \enquote{probability theory isn't true (quantum physics)}, and
purport to support these statements with the example of a quantum
double-slit experiment. Their statements are
false. % In the present note I recall some literature in
% quantum theory that shows why it is false, sum up the incorrect reasoning
% underlying their example, and also correct some wrong or imprecise
% statements about the quantum physics in that example.
In fact, opposite statements can be made, from two different
perspectives: %, which will be discussed in the rest of this note:
\begin{itemize}
\item The example given in that manuscript confirms, rather than
  invalidates, the probability rules. The probability calculus shows that a
  particular relation between probabilities, to be discussed below,
  \emph{cannot a priori} be assumed to be an equality or an inequality. In
  the quantum example it turns out to be an inequality, thus confirming
  what the probability calculus says.
\item But actually the same inequality can be shown to appear in very
  non-quantum examples, such as drawing from an urn. Thus there is nothing
  peculiar to quantum theory in this matter.
\end{itemize}

In the present comment I will prove the two points above, recalling some
relevant literature in quantum theory. I shall also correct a couple of
wrong or imprecise statements that Gelman \amp\ Yao make about quantum
physics in their example.

Let me point out at the outset that the rules of probability theory
(product or conditional or conjunction, sum or disjunction, negation) are
in fact routinely used in quantum theory with full validity, especially in
problems of state \enquote{retrodiction} and measurement reconstruction
\citep{jones1991b,slater1995b}[\chaps~7,8]{demuynck2002b_2004}{barnettetal2003,zimanetal2004_r2006,darianoetal2004}[see][\sect~1
and the rest of the present comment for many further
references]{maanssonetal2006}. An example is the inference of the state of
a quantum laser given its output through different optical apparatus
\citep{leonhardt1997}.

Similar incorrect claims with similar examples have appeared before in the
quantum literature \citep[see \eg][]{brukneretal2001}. Bernard O. Koopman
\footcite[of the Pitman-Koopman theorem for sufficient
statistics,][]{koopman1936} discussed the falsity of such claims already in
\cite*{koopman1957}. The Introduction in his work is very clear:
\begin{quotation}\footnotesize
  Ever since the advent of modern quantum mechanics in the late
  1920's, the idea has been prevalent that the classical laws of
  probability cease, in some sense, to be valid in the new theory. More or
  less explicit statements to this effect have been made in large number
  and by many of the most eminent workers in the new physics \textelp{}.
  Some authors have even gone farther and stated that the formal structure
  of logic must be altered to conform to the terms of reference of quantum
  physics \textelp{}.

  Such a thesis is surprising, to say the least, to anyone holding more or
  less conventional views regarding the positions of logic, probability,
  and experimental science: many of us have been apt -- perhaps too naively
  -- to assume that experiments can lead to conclusions only when worked up
  by means of logic and probability, whose laws seem to be on a different
  level from those of physical science.

  The primary object of this presentation is to show that the thesis in
  question is entirely without validity and is the product of a confused
  view of the laws of probability.
\end{quotation}

% A more recent claim, somewhat similar to Gelman \amp\ Yao's and with a
% similar supporting example, was made in a work by Brukner \amp\ Zeilinger
% \citey{brukneretal2001} and disproved by Porta Mana
% \citey{portamana2003_r2004} through a step-by-step analysis and
% calculation.
% ; although their focus was on an alleged
% inconsistency of some properties of the Shannon entropy in quantum
% theory.

It must be remarked that such claims have hitherto never been supported by
any rigorous proof -- with explicitly stated definitions and assumptions,
well-defined and unambiguous notation, and clear logical and mathematical
steps (and Gelman \amp\ Yao are no exception). The typical fallacy in the
kind of examples presented rests in the neglect of the experimental setup,
leading either to an incorrect calculation of conditional probabilities, or
to the incorrect claim that the probability calculus yields an equality,
where it actually does not. The same incorrect claims can be obtained
\emph{with completely non-quantum systems}, such as drawing from an urn, if
the setup is neglected
\citep{kirkpatrick2001_r2003,kirkpatrick2002_r2003}[\sect~IV]{portamana2003_r2004}.

Let us start with such a non-quantum counter-example.

\bigskip

\paragraph{A non-quantum counter-example}

Consider an urn with one $B$lue and one $R$ed ball. Two possible
drawing setups are given:
\begin{enumerate}%[label=$D_{\textrm{\alph*}}$:,ref=$D_{\textrm{\alph*}}$]
\item[$\yDa$]\label{item:repB} With replacement for blue, without replacement for
  red. That is, if blue is drawn, it is put back before the next draw (and
  the urn is shaken); if red is drawn, it is thrown away before the next
  draw.
\item[$\yDb$]\label{item:repR} With replacement for red, without replacement for
  blue.
\end{enumerate}
These two setups are obviously mutually exclusive.

We can easily find the unconditional probability for blue at the
second draw in the setup $\yDa$:
\begin{equation}
  \label{eq:A_red2}
  \p(B_{2} \| \yDa) = \tfrac{3}{4} \;.
\end{equation}
Note that this probability can be intuitively found by simple enumeration,
\`a la Boole, considering \enquote{possible worlds} if you like. Out of
four possible worlds, half of which has blue at the first draw, and the
other half has red, we can count that three worlds have blue at the second
draw.

The conditional probabilities for blue at the second draw,
given the first draw, are also easily found:
\begin{equation}
  \label{eq:A_red2cond}
  \p(B_{2} \| B_{1} \land \yDa) = \tfrac{1}{2} \qquad
  \p(B_{2} \| R_{1} \land \yDa) = 1 \;.
\end{equation}
We find that
\begin{equation}
  \label{eq:A_rule}
  \p(B_{2} \| \yDa) =%\mathrel{\bm{\ne}}
  \p(B_{2} \| B_{1} \land \yDa) \; \p( B_{1} \| \yDa) +
  \p(B_{2} \| R_{1} \land \yDa)  \; \p( R_{1} \| \yDa) \;,
\end{equation}
which is just the rule of conditional probability. It is in fact just the
systematization and generalization of the intuitive \enquote{possible
  worlds} reasoning done above.

\medskip

Next consider the setup $\yDb$. We easily find
\begin{gather}
  \label{eq:B_red2}
  \p(B_{2} \| \yDb) = \tfrac{1}{4} \;,
  \\
  \label{eq:B_red2cond}
  \p(B_{2} \| B_{1} \land \yDb) = 0 \qquad
  \p(B_{2} \| R_{1} \land \yDb) = \tfrac{1}{2} \;,
  \\[\jot]
  \label{eq:B_rule}
  \p(B_{2} \| \yDb) =%\mathrel{\bm{\ne}}
  \p(B_{2} \| B_{1} \land \yDb) \; \p( B_{1} \| \yDb) +
  \p(B_{2} \| R_{1} \land \yDb)  \; \p( R_{1} \| \yDb) \;.
\end{gather}

\medskip

Now compare the unconditional probability for blue at the second draw in
the setup $\yDa$, with the conditional probabilities for blue at the second
draw given the first draw in the setup $\yDb$:
\begin{equation}
  \label{eq:cross_combine}
  \p(B_{2} \| \yDa) \mathrel{\bm{\ne}}
  \p(B_{2} \| B_{1} \land \yDb) \; \p( B_{1} \| \yDb) +
  \p(B_{2} \| R_{1} \land \yDb)  \; \p( R_{1} \| \yDb) \;.
\end{equation}
This inequality is not surprising -- we are comparing different setups. It
is \emph{not} an instance of the conditional-probability rule. In fact the
probability calculus has nothing to say, a priori, about the relation
between the left side and right side, which are conditional on different
statements or, if you like, pertain to two different sample spaces.

You can call the inequality above \enquote{interference} if you want; for
further and more involved examples with urns and decks of cards see
Kirkpatrick \citey{kirkpatrick2001_r2003,kirkpatrick2002_r2003} and Porta
Mana \citey[\sect~IV]{portamana2003_r2004}.

\medskip

Now consider another pair of drawing setups: setup $\yDc$, with replacement
for both colours; and setup $\yDd$, without replacement for either colour.
You can easily find that
\begin{gather}
  \p(B_{2} \| \yDc) =%\mathrel{\bm{\ne}}
  \p(B_{2} \| B_{1} \land \yDc) \; \p( B_{1} \| \yDc) +
  \p(B_{2} \| R_{1} \land \yDc)  \; \p( R_{1} \| \yDc) \;,
  \\
  \p(B_{2} \| \yDd) =%\mathrel{\bm{\ne}}
  \p(B_{2} \| B_{1} \land \yDd) \; \p( B_{1} \| \yDd) +
  \p(B_{2} \| R_{1} \land \yDd)  \; \p( R_{1} \| \yDd) \;,
  \\
  \label{eq:cross_combine_same}
  \p(B_{2} \| \yDc) \mathrel{\bm{=}}
  \p(B_{2} \| B_{1} \land \yDd) \; \p( B_{1} \| \yDd) +
  \p(B_{2} \| R_{1} \land \yDd)  \; \p( R_{1} \| \yDd) \;.
\end{gather}
The first two equalities above are expressions of the
conditional-probability rule. The third is \emph{not}, however. It is
simply a peculiar equality contingent on the two specific setups.

The probability calculus therefore correctly handles situations leading to
inequalities such as~\eqref{eq:cross_combine}, and to equalities such
as~\eqref{eq:cross_combine_same}.

\medskip

% Strictly speaking it is wrong to use the expression \defquote{$B_{2}$} for
% all these setups, because \defquote{$B_{2}$} in each setup denotes a
% different statement (or random variable) than in the others. Just like
% \enquote{it rains (on 14 July 2020 in Trondheim)} is different from
% \enquote{it rains (on 20 November 2019 in Rome)}. I should have used
% different symbols.

The explicit presence of \defquote{$D_{\dotso}$}, which represents given
information, is necessary discussions involving different setups, such as
the above. If I ask you \enquote{what's the probability of blue at the
  second draw?}, you will ask me \enquote{in which drawing setup?}. The
probability is conditional on the information about the drawing scheme.

% luckily avoided any ambiguities. But if in our formulae we omit the
% notation of the setup \emph{and} we use the same notation for actually
% different statements or random variables, then we're in for trouble and for
% incorrect applications of the probability rules.

\medskip

The inequality~\eqref{eq:cross_combine} is what Gelman \amp\ Yao
\citey[p.~2]{gelmanetal2020} complain about, but in the context of a pair
of quantum setups. I do not see how one can complain about it, or claim
inconsistencies. It is obviously correct even from an intuitive analysis of
the two setups. And the probability calculus correctly leads to it, too.
The probability calculus correctly leads also to the
equality~\eqref{eq:cross_combine_same}. As already said, given two mutually
exclusive setups, the probability calculus a priori neither commits to an
equality nor to an inequality.

I will now show that the simple example above is in fact conceptually quite
close to the quantum experiment mentioned by Gelman \amp\ Yao. The
closeness is especially clear from the experimental and mathematical
developments of quantum theory of the past 40 years (at the very least), as
the literature cited below shows.

\bigskip

\paragraph{The quantum double-slit experiments}

The basic argument of Gelman \amp\ Yao is that, in a given setup of the
quantum double-slit experiment, we have a specific probability distribution
for the appearance of an emulsion or excitation on some point of the
screen. We can call this a \enquote{screen detection}, but please keep in
mind that in so doing we are adding an extra interpretation that modern
quantum theory does not actually commit to (see discussion and references
below). In a different experimental setup we have conditional probabilities
for screen detection conditional on slit detection. Now, the probability of
the first setup is not equal to the combination of the conditional
probabilities of the second setup.

But this is exactly what happened in our urn example above,
\eqn~\eqref{eq:cross_combine}. In the present quantum case we do \emph{not}
have a violation of the conditional probability rule either -- if anything
it is a confirmation.

To see the analogy more clearly, let me present some additional facts from
quantum theory.

\medskip

The experimental setup without detectors at the slits and the setup with
slit detectors are actually limit cases of a continuum of experimental
setups \citep{woottersetal1979}[for a recent review and further references
see][]{banaszeketal2013}. In the general case, such a setup has slit
detectors of varying efficiency, denoted by a parameter $q \in \clcl{0, 1}$
that can be chosen in the setup. The possible degrees of efficiency are of
course mutually exclusive, so these setups are mutually exclusive.

The slit detector has a given efficiency in the following sense:

Let us call $y$ the detection position on the screen, and $\yxa$ is the
statement that detection occurs at slit \#1 (you can translate to
random-variable jargon if you prefer). When we prepare the electromagnetic
field in a quantum state $S$, and use ideal detectors with perfect
efficiency, the probability of detection at slit \#1 is, say $p_{S}$, and
$1-p_{S}$ for slit \#2.

% in such a way that, in the setup with
% non-noisy slit detectors, detector \#1 always fires; that is, statement
% $\yxa$ has probability 1 and statement $\yxb$ has probability 0 in this
% setup. We can also prepare a state $S_{2}$ such that the opposite holds in
% the same non-noisy setup: $\yxa$ has zero probability, and $\yxb$ unit
% probability.

If we use the setup with detectors having efficiency $q$ -- denote it by
$\yDq$ -- then the probability of detection at slit \#1 is
% n we have probability $q$ that slit detector \#1 fires, when
% state $S_{1}$ is prepared; and probability $q$ that slit detector \#2
% fires, when state $S_{2}$ is prepared; the probabilities for the other slit
% are $1-q$. That is,
\begin{equation}
  \label{eq:explain_noisy}
  \pf(\yxa \| \yDq, S) =  \tfrac{1}{2} (1-q) + q\,p_{S}\;,
\end{equation}
and $\tfrac{1}{2}(1+q) - q\,p_{S}$ for slit \#2.
% (We should actually not be using the same $X$ symbol for all such setups,
% because the sample spaces are different; see the discussion for the urn
% example).

The setup with perfect detectors is the limit case $q=1$. In the case of
zero efficiency, $q=0$, there is no relation between the light states and
the firing of the slit detectors; that is, we are always fully uncertain as
to which detector would fire, no matter how the light state is prepared.
These kinds of setup -- and many other interesting ones -- are quite easy
to prepare with the statistically analogous quantum Mach-Zehnder-like
interferometers (see the textbooks in footnote~\ref{fn:ref_interf} below;
\cites[\sect~4.2]{leonhardt1997}{yuenetal1978}).
  
In each setup $\yDq$ (and given the light state $S$) we also have the
conditional probability distribution $\pf(y \| \yDq,S)$ for detection at
$y$ on the screen, and the conditional probability distributions
$\pf(y \| X,\yDq,S)$ for detection at $y$ on the screen, given detection
$X$ at the slits. We have
\begin{multline}
  \label{eq:conditional_q}
  \pf(y \| \yDq, S) ={}\\
  \pf(y \| \yxa, \yDq, S) \;\pf( \yxa \| \yDq, S) +
  \pf(y \| \yxb, \yDq, S) \;\pf( \yxb \| \yDq, S) \;.
\end{multline}
This is an instance of the conditional-probability rule, which is of course
valid. This equality also holds for long-run frequencies (see point
\ref{item:q_prob_freq} below). Note that such conditional and unconditional
frequencies are experimentally observed. I would like you to convince
yourself, though, that the \emph{equality} above (not the specific values
of the frequencies) is not really an experimental fact, since it rests on
the very way we measure conditional frequencies.

The conditional and unconditional distributions above will of course be
different depending on the setup $\yDq$ and the light state $S$. But in
each instance the rule of conditional probability holds. For example, if
$\yqa \ne \yqb$,
\begin{gather}
  \begin{multlined}[][\linewidth]
    \pf(y \| \yDqa, S) ={}\\
  \pf(y \| \yxa, \yDqa, S) \;\pf( \yxa \| \yDqa, S) +
  \pf(y \| \yxb, \yDqa, S) \;\pf( \yxb \| \yDqa, S) \;,
\end{multlined}
\\
  \begin{multlined}[][\linewidth]
    \pf(y \| \yDqb, S) ={}\\
  \pf(y \| \yxa, \yDqb, S) \;\pf( \yxa \| \yDqb, S) +
  \pf(y \| \yxb, \yDqb, S) \;\pf( \yxb \| \yDqb, S) \;,
\end{multlined}
\\
  \begin{multlined}[][\linewidth]
    \pf(y \| \yDqa, S) \mathrel{\bm{\ne}} {}\\
  \pf(y \| \yxa, \yDqb, S) \;\pf( \yxa \| \yDqb, S) +
  \pf(y \| \yxb, \yDqb, S) \;\pf( \yxb \| \yDqb, S) \;.
\end{multlined}
\end{gather}
The last inequality, analogous to \eqn~\eqref{eq:cross_combine}, comes from
experimental observations (see the brief discussion below about the
relation with de~Finetti's theorem), and was not in fact not ruled out a
priori by the probability calculus.

\medskip

Now let me discuss a couple of very interesting experimental facts about
this collection of setups:

First, \emph{both the conditional $y\vert X$ and unconditional} probability
distributions for the screen detection $y$ generally \emph{have an
  oscillatory profile, typical of interference}
\citep{woottersetal1979,banaszeketal2013}[see also][for other experimental
variations]{chiaoetal1995}. The oscillatory character is maximal for the
zero-efficiency setup $q=0$ and decreases as $q$ increases. For the
perfect-detector setup $q=1$ there is no interference. But we can have
quite a lot of interference even when the detection efficiency is quite
high, so that for some light states we are almost certain about slit
detection; see references above. (The profile depends on the specific light
state, of course, which we are assuming fixed.)
  
Second, the unconditional (frequency) distribution observed in the setup
$D_{0}$ with zero-efficiency slit detectors is experimentally equal to the
distribution for screen detection observed in the setup without slit
detectors (note that in the latter setup we cannot speak of conditional or
unconditional probability, since slit detection does not exist).

Third, one conditional distribution observed in the setup with one slit
closed is experimentally equal to one in the setup $D_{1}$ with perfect
slit detectors. (Here we must be careful, because there is no slit
detection in the second setup; rather, we speak of appearance or
non-appearance at the screen, and in the latter case no conditional
distribution is defined.)

The equalities in the last two cases should a priori not be expected,
because the setups are physically different. Of course one can look for
physical, \enquote{hidden variables} explanations of such equalities.
Experimental quantum optics simply acknowledges the fact that two setups
are equivalent for such detection purposes, and incorporates this
information into its mathematical formalism, by means of appropriately
defined \defquote{\povm s}, discussed below.

\medskip

Note the statistical analogy between the cases above and the cases with the
setups of the urn examples previously discussed. In each setup, the rule of
conditional probability holds (and in the quantum case we can have
distributions, conditional and unconditional, with oscillatory profiles).
Across different setups, probability theory says that such a rule cannot be
applied; and indeed we find inequalities across some setups and equalities
across others, both in the quantum and non-quantum case,
\eqns~\eqref{eq:cross_combine}, \eqref{eq:cross_combine_same}. Even more
striking statistical analogies appear in the already cited non-quantum
counter-examples
\citep{kirkpatrick2001_r2003,kirkpatrick2002_r2003}[\sect~IV]{portamana2003_r2004}.

\medskip

It is also possible to consider situations in which we are uncertain about
which measurement setup applies. For example we may not know whether there
were slit detectors, or the value $q$ of the detector efficiency. In such
situations we introduce probabilities $\pf(D_{\dotso})$ for the possible
setups and the conditional-probability rule applies, yielding for example
\begin{equation}
  \label{eq:conditional_setups}
  \pf(y \| S) = \tsum_{q} \pf(y \| \yDq, S)\;\pf(\yDq)
\end{equation}
(here our knowledge of the state was assumed to be irrelevant to our
inference about the setup). Then, given the measurement outcome, we can
make inferences about the setup
\citep{barnettetal2003,zimanetal2004_r2006,darianoetal2004}[see
also][]{rigoetal1998} -- for example whether a slit detector was present or
not -- again using the conditional-probability rule in the guise of Bayes's
theorem. This kind of inference is especially important in quantum key
distribution \citep{nielsenetal2000_r2010}, where we try to infer whether a
third party was eavesdropping, that is, performing a covert measurement.
Again no violations of the probability rules in the quantum realm: quite
the opposite, those rules allow us to make important inferences.

\bigskip

\paragraph{Further remarks and curiosities about quantum interference
  experiments}

I would like to mention a couple more experimental facts -- which are,
besides, statistically very interesting -- to correct some statements by
Gelman \amp\ Yao in relation to the double-slit experiment.
% discussion the experiment further, also mentioning latent variables, but
% their discussion is misleading and incorrect from some points of view.
% Therefore I would like to remark on some very interesting facts and provide
% literature about this quantum experiment.

\begin{enumerate}[label=(\textbf{\roman*})]
%\item\label{item:continnum_experiments} 
\item\label{item:q_details} It \emph{does} matter whether many photons are
  sent at once, or one at a time \citep[\cf][\sect~2
  point~1]{gelmanetal2020}; as well as their wavelength, temporal spread,
  and so on (strictly speaking, the spatio-temporal dependence of the field
  mode). These details are part of the specification of the light state $S$
  mentioned above, and lead to different probabilities distributions of
  screen detection.

  For example, in some setups and for some states we can have a detection
  probability density $\pf(y_{1})$ for the first photon, and a
  \emph{different} density for the second photon $\pf(y_{2} \| y_{1})$,
  conditional on the detection of the first -- both being different from
  the cumulative density of detections. Interference phenomena can also be
  observed in time, not only in space. See \eg\ the phenomena of
  higher-order coherence, bunching, anti-bunching, and many other
  interesting ones\footnote{\label{fn:ref_interf}\cites[for
    example][]{mandeletal1965,morganetal1966,paul1982,jacobsonetal1995}[and
    textbooks such
    as][]{loudon1973_r2000,mandeletal1995_r2008,scullyetal1997_r2001,gerryetal2005,wallsetal1994}.}.
  Quoting Glauber \citey[Lect.~I p.~65]{glauber1965}:
  \begin{quote}\footnotesize
    The new light detectors enable us to ask more subtle questions than
    just ones about average intensities; we can for example, ask questions
    about the counting of pairs of quanta, and can make measurements of the
    probability that the quanta are present at an arbitrary pair of space
    points, at an arbitrary pair of limes.
  \end{quote}
  The rules of the probability calculus also apply in all such situations.
  We can infer, for example, the position of the first photon detection
  given the second from the conditional probability rule
  $\pf(y_{1} \| y_{2}) \propto \pf(y_{2} \| y_{1})\;\pf(y_{1})$.

\item\label{item:latent_vars} The details about the light source and the
  setup are not \enquote{latent variables}: they specify the quantum state
  of light and the measurement performed on it. They are like the initial
  and boundary conditions necessary for the specification of the behaviour
  of any physical system. % They correspond to the different drawing
  % setups in the urn example above -- we would not call the specification
  % \enquote{drawing with replacement} a latent variable. The rules of
  % probability apply seamlessly in each case.

\item\label{item:q_prob_freq} In view of point~\ref{item:q_details} above,
  it is important not to conflate the probability distributions for
  single-photon detections, those for cumulative photon detection, and the
  \emph{frequency} distributions of a long-run of such detections
  \citep[\sect~2, seem to conflate the two]{gelmanetal2020}. Such
  distinction is always important from a Bayesian point of view. Quoting
  Glauber \citey[Lect.~I p.~70]{glauber1965} again:
  \begin{quote}\footnotesize
    There is therefore no possibility \emph{in general} of replacing the
    ensemble averages by time averages. \textelp{} we shall find that
    individual measurements yield results wholly unlike their ensemble
    averages. The distinction between particular measurements and their
    averages may thus be quite essential.
  \end{quote}
\end{enumerate}

I may add that the idea and parlance of \enquote{photons passing through
  slits} are used today only out of tradition; maybe a little poetically.
The technical parlance, as routinely used in quantum-optics labs for
example \citep{leonhardt1997,gerryetal2005}, has a different
underlying picture. The \defquote{system} in a quantum-optics experiment is
not photons, but the modes of the field-configuration operator\ibid\ (note
that this is not yet Quantum ElectroDynamics). \enquote{Photon numbers}
denote the discrete outcomes of a specific energy-measurement operator;
\enquote{photon states} denote specific states of the field operators. As
another example, \enquote{entanglement} is strictly speaking not among
photons, but among modes of the field operator \citep{vanenk2003b}. Several
quantum physicists indeed oppose the idea and parlance of
\enquote{photons}, owing to the confusion they lead to. Lamb \footcite[of the
Lamb shift,][]{lambetal1947} wrote in \cite*{lamb1995}:
\begin{quote}
  \footnotesize the author does not like the use of the word ``photon'',
  which dates from 1926. In his view, there is no such thing as a photon.
  Only a comedy of errors and historical accidents led to its popularity
  among physicists and optical scientists.
\end{quote}
Wald \citey{wald1994} warns:
\begin{quote}
  \footnotesize standard treatments of quantum field theory in flat
  spacetime rely heavily on Poincar\'e symmetry (usually entering the
  analysis implicitly via plane-wave expansions) and interpret the theory
  primarily in terms of a notion of ``particles''. Neither Poincar\'e (or
  other) symmetry nor a useful notion of ``particles'' exists in a general,
  curved spacetime, so a number of the familiar tools and concepts of field
  theory must be ``unlearned'' in order to have a clear grasp of quantum
  field theory in curved spacetime. [p.~ix] \textelp{} the notion of
  ``particles'' plays no fundamental role either in the formulation or
  interpretation of the theory. [p.~2]
\end{quote}
See also Davies's \emph{Particles do not exist} \citey{davies1984}.

\bigskip

\paragraph{A summary of the modern formalism of quantum theory}

It may be useful to give a summary of how probability enters the modern
formalism of quantum theory. See textbooks such as Holevo
\citey{holevo1980_t2011}, Busch \etal\ \citey{buschetal1995b}, Peres
\citey[especially \chap~12]{peres1993_r2002}, de Muynck \citey[especially
\chap~3]{demuynck2002b_2004}, and the excellent text by Bengtsson \amp\
\.Zyczkowski \citey{bengtssonetal2006_r2017}.

A quantum system is defined by its sets of possible states and possible
measurements. A state $\rho$ is represented by an Hermitean,
positive-definite, unit-trace matrix $\yr$ \citep[which satisfies
additional mathematical
properties:][]{jakobczyketal2001,kimura2003,kimuraetal2004,bengtssonetal2006_r2017},
called \defquote{density matrix}. States traditionally represented by kets
$\lvert \psi \rangle$ are just special cases of density matrices. A
measurement setup $M$ is represented by a set of Hermitean,
positive-definite matrices $\set{\yM_{\ybr}}$ (of the same order as the
density matrices) adding up to the identity matrix. They are called
\defquote{positive-operator-valued measures}, usually abbreviated \povm s.
Traditional von~Neumann projection operators
$\set{\lvert \phi_{r} \rangle\langle \phi_{r}\rvert}$ are just special
cases of \povm s. Each matrix $\yM_{\ybr}$ is associated with an outcome
$\ybr$ of the measurement. These outcomes are mutually exclusive. An
outcome can actually represent a combination of simpler outcomes,
$\ybr\equiv(x,y,z,\dotsc)$, such as the intensities or firings at two or
more detectors.

The probability of observing outcome $\ybr\equiv (x, y, \dotsc)$ given
the measurement setup $M$ and the state $S$ is encoded in the trace-product
of the respective matrices:
\begin{equation}
  \label{eq:povm_trace}
  \pf(x,y,\dotsc \| M \land S) \equiv \tr(\yM_{x,y,\dotsc}\;\yr) \;,
\end{equation}
These probabilities for all $\ybr$ form a probability distribution. The
traditional Born-rule expression
\defquote{$\abs{\langle \phi_{r} \vert \psi \rangle}^{2}$} is just a
special case of the above formula. The probabilities in the formula come
from repeated measurement observations in the same experimental conditions:
we can invoke de~Finetti's \citey{definetti1937,definetti1938} theorem here
-- the partial-exchangeability variant -- and some quantum physicists
indeed do \citep{cavesetal2002,vanenketal2002,fuchsetal2004b}. The
trace-product above is just a scalar product in a particular space. How a
set of probability or frequency distributions can be encoded in scalar
products is explained in a down-to-earth way in Porta Mana
\citey{portamana2003b,portamana2004b}.

Once the probability distribution above is given we can use the
full-fledged probability calculus for our inferences. We can for example
sum (or integrate) over detector outcomes $y,\dotsc{}$, obtaining the
marginal probability for detector outcome $x$; or calculate the probability
of outcome $y$ conditional on $x$; or make inferences about the measurement
setup or the state. Again, there are no violations of the probability
rules. The formalism~\eqref{eq:povm_trace} is neat in this respect because
it allows us to represent such situations through new \povm s and density
matrices. You can easily check, for example, that the marginal probability
for $x$ from \eqn~\eqref{eq:povm_trace} can be encoded in the \povm\
$\set{\yM'_{x}} \equiv \set{\tsum_{y,\dotsc}\yM_{x,y,\dotsc}}$. A situation
of uncertainty between setups $M'$ and $M''$, as in
\eqn~\eqref{eq:conditional_setups}, can be encoded in the \povm\
$\set[\big]{\pf(M')\;\yM'_{\ybr} + \pf(M'')\;\yM''_{\ybr}}$. And so on, and
similarly for states and their density operators.

For systems with infinite degrees of freedom such as electromagnetic fields
or electrons (Fermionic fields), the matrices above are replaced by
operators defined in particular algebras. A \povm\ element can actually be
a space-time-indexed operator. The computational details can become quite
complicated, but the same basic ideas apply.

This formalism obviously also includes the specification of
post-measurement states (if the system still exists afterwards),
transformations, evolutions. I shall not discuss these; see the textbooks
cited above.

\bigskip

\paragraph{Conclusions}

I hope that the above discussion and bibliography clearly show that:
\begin{itemize}
\item the rules of probability theory, including the
  conditional-probability rule, are fully valid in quantum theory and
  essential in its modern applications;
\item some peculiar equalities or inequalities across different
  experimental conditions do not contradict the conditional-probability
  rule, and they appear just as well in quantum as in non-quantum
  situations, such as drawing from an urn.
\end{itemize}

Quantum theory already has its physically conceptual difficulties and
computational difficulties, as should be clear from the portrait sketched
in the present comment. It is pointless -- and pedagogically confusing and
detrimental, for students of quantum optics for instance -- to make it seem
even more difficult with false claims of non-validity of probability theory
or with distorted pictures of its experimental content.

\bigskip
% %\renewcommand*{\appendixpagename}{Appendix}
% %\renewcommand*{\appendixname}{Appendix}
% %\appendixpage
% \appendix

%%%%%%%%%%%%%%%%%%%%%%%%%%%%%%%%%%%%%%%%%%%%%%%%%%%%%%%%%%%%%%%%%%%%%%%%%%%%
%%% Bibliography
%%%%%%%%%%%%%%%%%%%%%%%%%%%%%%%%%%%%%%%%%%%%%%%%%%%%%%%%%%%%%%%%%%%%%%%%%%%% 
\renewcommand*{\finalnamedelim}{\addcomma\space}
\defbibnote{prenote}{{\footnotesize (\enquote{de $X$} is listed under D,
    \enquote{van $X$} under V, and so on, regardless of national
    conventions.)\par}}
% \defbibnote{postnote}{\par\medskip\noindent{\footnotesize% Note:
%     \arxivp \mparcp \philscip \biorxivp}}

\printbibliography[prenote=prenote%,postnote=postnote
]

\end{document}